AC losses in superconducting composite strips in a magnetic field in the form of a standing wave

V. Sokolovsky, V. Meerovich,

Physics Department, Ben-Gurion University of the Negev, P. O. Box 653, Beer-Sheva 84105, Israel

**Abstract**

Analytical expressions for the evaluation of AC losses in a superconductor-metal composite strip in a nonuniform AC magnetic field having the form of a standing wave are derived. The considered configuration models superconducting tapes, thin films and coated superconductors. A distinctive feature of the problem is the appearance of a transverse component of the induced current that converts the problem to a two-dimensional one. Bean's critical state model for a superconductor and a low frequency approximation for a metal are used. In the framework of this approximation the influence of eddy currents in the metal on the magnetic field is neglected, and the current distribution in the superconductor is determined by an external field. Two cases were considered: (1) superconducting and metal strips electrically separated; (2) they are in an electric contact. It is shown that for most practical cases enough to take into account only the losses generated by induced longitudinal currents in a composite strip. When an electric contact between strips exists, the maximum loss density in high fields (above 2 T) can be determined by the transverse component of the current.

**1. Introduction**

Platelet geometry is typical for many superconducting products: tapes, thin films, coated superconductors where a superconducting strip is connected parallel with a normal conducting strip or confined in a normal conducting matrix. Theoretical models based on the consideration of thin and infinitely long metal and superconducting strips placed one on the top of the other were developed for calculation of magnetization and losses in superconducting composites in a uniform AC magnetic field [1-9]. In many practical situations, however, the external field is not exactly uniform and can be better presented as a superposition of a uniform part and spatiotemporal fluctuations along a conductor. In the case of flywheel system, generators and motors the fluctuations are induced by rotation of permanent magnets and may be approximated by a running wave. In the case of magnetic shields or transformers a nonuniform magnetic field in the form of a standing wave can serve as a better approximation.

In this paper we evaluate AC losses in a thin metal-superconductor strip in a nonuniform perpendicular magnetic field having the form of a standing wave along the strip. A distinctive feature of the problem is the appearance of the lateral component of the induced current (so called "closure" currents) with the result that the problem becomes two-dimensional. Following the researchers studying the case of a uniform field [8, 9], we use Bean's critical state model for a superconducting strip and a low frequency approximation for a metal strip. In the framework of this approach the influence of eddy currents in the metal on the magnetic field at the composite can be neglected, and the current distribution in the superconductor is determined by an external field. Afterwards, the metal strip is considered in the field which is the sum of the external one and the field generated by the current in the superconductor. We analyze the influence of the contact between superconducting and normal strips on the losses in a composite.

## 2. Hysteresis losses in a superconducting strip

Let us consider a thin type-II superconducting strip, infinitely long along the *x*-axis, with the width 2*w* along *y* and the thickness *a* along *z*. The strip is placed in an external field in the form of a standing wave which does not possess a component directed along the strip ($H_x = 0$):

$$\vec{H} = H_0 \cos(\omega t) \left[ 0, \sin(\varphi) \cos(kx), \cos(\varphi) \cos(kx) \right] \qquad (1)$$

where $H_0$ and $\omega = 2\pi f$ is the amplitude and cyclic frequency of the external magnetic field, respectively, $\varphi$ is the angle between the vectors of the magnetic field and of the normal to the film surface, $k = 2\pi/L$, $L$ is the length of the standing wave.

The magnetic field is directed under an arbitrary angle to the film wide-face and what is more the angle can change down the strip. Since $H_x = 0$, the task is reduced to 2D-problem for the induced currents which have the distribution schematically presented in Fig. 1. The distribution is repeated with the period of a half-wave of the applied magnetic field $L/2$.

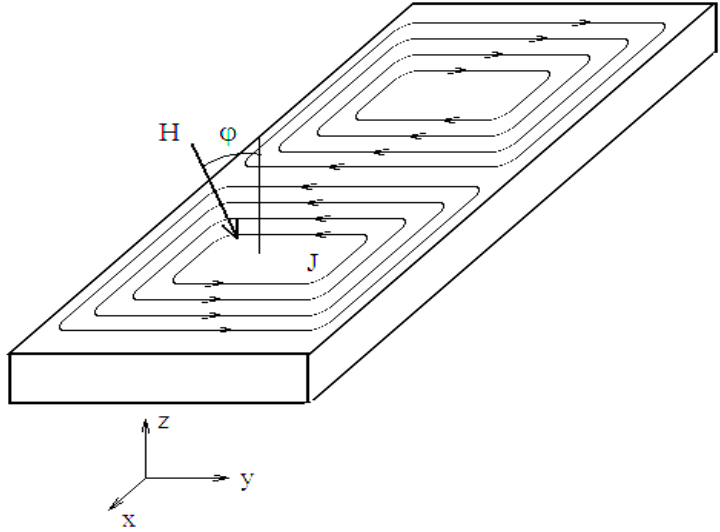

Fig. 1. Qualitative distribution of induced currents in a strip in a nonuniform magnetic field.

The following qualitative analysis shows that one can neglect by a loss part due to the *y*-component of the magnetic field. Actually, both components of the magnetic field induce the *x*-component of the electric field. The part of the electric field induced by the *z*-component $H_z$ of the magnetic filed can be estimated as $E_{xz} \sim \omega w(y/w)H_z$ while the part of the electric field induced by the *y*-component $H_y$ is $E_{xy} \sim \omega a(z/a)H_y$. Here $0 \leq y/w \leq 1$ and $0 \leq z/a \leq 1$. The direction of the current density coincides with the *E* direction in a homogeneous superconductor and equals to the critical value $j_c$ in the critical state model. The ratio of the loss per a unit of the strip length from the *y*- and *z*-components of the magnetic field is proportional to $(a/w)^2$. The thickness of the superconducting layers in films, tapes and coated conductors is the order of few micrometers or less while their width is several millimeters. Therefore, the contribution of the *y*-component in the losses is negligible (except the case when the applied field is close to parallel). The problem is reduced to a strip in a perpendicular external field:

$$\vec{H} = H_0 \cos(\omega t)[0,0,\cos(kx)] \qquad (2)$$

The magnetic field from the induced currents is a sum of the fields produced by the *x*- and *y*-current components. Since $L >> w$, at least within the accuracy of $w/L$, the contribution of the *y*-component (closure currents) in the total magnetic field can be neglected in the comparison with the contribution of the *x*-component. Therefore, the distribution of the *x*-component of the currents is determined only by a local applied field *H*. To calculate this distribution, we use the expression obtained using the Bean model in [1, 2] for the sheet density $J_x$ of the current in a thin strip in a uniform perpendicular magnetic field, monotonically increasing from 0 to *H* :

$$J_x(y, H, J_c) = \begin{cases} \dfrac{2J_c}{\pi} \text{acrtan}\left[\dfrac{cy}{\sqrt{b^2 - y^2}}\right], & |y| < b \\ J_c y / |y|, & b < |y| < w \end{cases} \qquad (3)$$

where $b = w/\cosh(H/H_c)$, $c = \tanh(H/H_c)$, $H_c = J_c/\pi$, $J_c = 2aj_c$ is the sheet critical current density.

When the applied field oscillates between the values $\pm H_a$ with the angular frequency $\omega$, Eq. (3) is used to find the instant distribution of the current. For example, considering a decrease of the applied field from $+ H_a$ to $- H_a$, one can obtain [1]:

$$J_{dx}(y, H_2, J_c) = J_x(y, H_a, J_c) - J_x(y, H_a[1 - \cos(\omega t)], 2J_c). \qquad (4)$$

To determine the distribution of the $x$-component of the current in the considered case of a nonuniform field, we replace $H_a$ by its function of $x$: $H_a = H_0 \cos(kx)$. In this case, $c$ and $b$ in Eq. (3) become functions of the $x$-coordinate, and $J_{dx}$ turns a function of $x$ and $y$.

The $y$-component of the current can be estimated as

$$J_y(y) \quad \int_{-w}^{y} \frac{\partial J_x}{\partial x} dy . \qquad (5)$$

Note that Eq. (4) contains two parts: the first gives a current induced by the maximum of the magnetic field $H_a = H_0\cos(kx)$, the second is associated to a current appearing with the decrease of the field to $H = H_0\cos(kx)\cos(\omega t)$. The first part is equal to the critical sheet currents $\pm J_c$ in the penetration regions $|y| > b_1$, where $b_1 = w/\cosh(h\cos(kx))$, $h = H_0/H_c$. Similarly, the second part is equal to $\pm J_c$ in the regions $|y| > b_2$, where $b_2 = w/\cosh\{h\cos(kx)[1 - \cos(\omega t)]/2\}$. In these ranges the derivative with respect to $x$ from the corresponding part is zero. Therefore Eq. (5) can be presented in the following form

$$J_{dy}(y) = \int_{-b_1}^{y} \frac{dJ_x(y, H_0\cos(kx), J_c)}{dx} dy - \int_{-b_2}^{y} \frac{dJ_x(y, H_0\cos(kx)[1 - \cos(\omega t)], 2J_c)}{dx} dy . \qquad (6)$$

The first integral in Eq. (6) gives

$$J_{dy1} = \frac{\sqrt{2}J_c hkw}{\pi} \frac{\sin(kx)}{\cosh[h\cos(kx)]} \sqrt{2 - \frac{y^2}{w^2}\{1 + \cosh[2h\cos(kx)]\}} . \qquad (7)$$

The second integral $J_{dy2}$ in Eq. (6) can be obtained from Eq. (7) by replacing $h$ by $h[1 - \cos(\omega t)]/2$ and $J_c$ by $2J_c$. The obtained expressions for the integrals are valid: $J_{dy1}$ for $-b_1 < y < b_1$ and $J_{dy2}$ for $-b_2 < y < b_2$. Out of these ranges these integrals equal to zero.

Note that $J_{dy1}$ cannot be higher than $J_c$ and, therefore, one of the limitations of the proposed approach is

$$\frac{2hkw}{\pi} \le 1, \text{ or } \quad h \le h_l = \frac{\pi}{2kw}. \tag{8}$$

The loss power induced by the magnetic field in a conductor with zero transport current can be calculated using the following expression [10, 11]:

$$P = -f \int_V dV \int_0^T \vec{j} \cdot \frac{\partial \vec{A}_e}{\partial t} \tag{9}$$

where $j$ is the current density induced in the conductor, $A_e$ is the vector potential of the external magnetic field, $T$ is the period, $V$ is the volume of the conductor.

Generally, the vector potential is determined by the distribution of the currents producing an external magnetic field. Assuming the spatial symmetry of the problem we can obtain the vector potential as a periodical along the $x$-coordinate solution of the equations $\text{rot}\vec{A}_e = \vec{B}$ and $\text{div}\vec{A}_e = 0$ (see Appendix). In a first approximation at $z = 0$ the vector potential is

$$\vec{A}_e = \mu_0 H_0 \cos(\omega t)\left[-y\cos(kx), \ -\frac{ky^2}{2}\sin(kx), \ 0\right] \tag{10}$$

where $\mu_0$ is the magnetic permeability of vacuum.

The total losses can be presented as a sum of two parts. The first part is associated with the loss from the $x$-component of the current density $J_{dx}$. The loss power per unit length of the strip (loss density) is

$$p_x(h,x) = 2\mu_0\omega w^2 J_c H_c F_x(h,x)/\pi = P_0 F_x(h,x) \tag{11}$$

where $P_0 = 2\mu_0\omega w^2 J_c H_c/\pi$, $F_x(h,x) = 2\ln\left[\cosh(Z)\right] - Z\tanh(Z)$, $Z = h\cos(kx)$.

This differs from the known expression for the losses in a uniform field [1] only by the dependence of the magnitude $H_a$ on the coordinate.

The loss density related to the $y$-component of the current density $J_{dy}$ is

$$p_y(h,x) = (kw)^2 P_0 F_y(h,x) \tag{12}$$

where

$$F_y(h,x) = \frac{\sin(kx)^2}{24}\{\text{sech}(\cos(Z))^2 \sec(kx)\left[\sec(kx) + 2h\left[2 + \cos(Z)\right]\right]\tanh(Z) - 3h^2\text{sech}(\cos(Z))^4 - \left[1 + 4\ln(\cosh(Z))\right]\sec(kx)^2\} \ . \tag{13}$$

The dependence of the losses on the magnetic field and coordinate is determined by the functions $F_x$ and $F_y$ given by Eqs. (11) and (13). These functions are shown in Fig. 2 for $-\pi/2 < kx < \pi/2$. Due to symmetry of the task, the functions are periodic with respect $x$ with the period $L/2$. The maximum values of these functions (and, hence, maximums of the losses) are observed at different points: while the maximum of $F_x$ is always at $x = 0$; $F_y$ achieves the maximums about at the points $kx = \pm \pi/2$, at that the maximums approach to these points when the applied field increases. Fig. 3 demonstrates the behavior of the maximums vs. the magnetic field. The ratio of the maximum losses $\max(p_y)/\max(p_x)$ is proportional to $(kw)^2$. Analysis of the expressions (11)-(13) shows that for $h \ll 1$ the maximums of losses $p_x$ and $p_y$ are proportional to $h^4$. For $1 \ll h < h_l$, $\max(p_x)$ is proportional to $h$ and $\max(p_y)$- to $h^2$. More accurate analysis of the expressions for $p_x$ and $p_y$ for $h \gg 1$ gives the estimate:

$$\max(p_y)/\max(p_x) = (kw)^2 h/40. \qquad (14)$$

Taking into account the condition (8), the maximum of this ratio should be $w/4L$. Therefore, when $L \gg w$, the maximum of $p_x$ is sufficiently higher than that of $p_y$.

The local losses given by Eqs. (11) and (12) are important at design of a superconducting device because they determine the places where more intense cooling should be provided. The second important parameter related to the required power of the cooling equipment is the average losses per unit length of the conductor. They are determined by

$$< P > = \frac{4}{L} \int_{0}^{L/4} p(x,h)dx \quad . \qquad (15)$$

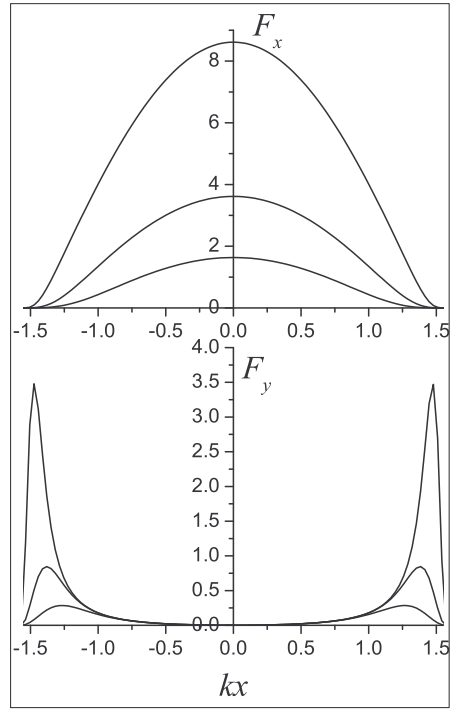

Fig. 2. Dependence of the functions $F_x$ and $F_y$ on the coordinate along a strip for different magnetic fields.

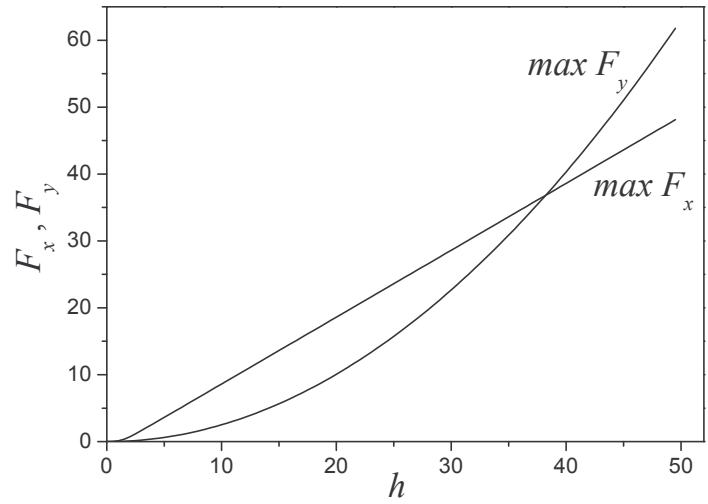

Fig. 3. Dependence of the maximum values of the functions $F_x$ and $F_y$ on the normalized magnetic field $h$.

The contribution of the $y$-component in the average losses is determined by the ratio of the averaged functions (Fig.4) multiplied by the coefficient $(kw)^2$. At $L >> w$ and $h < h_l$ the total losses are determined practically by the $x$-component. Fig. 5 shows that the ratio of the maximum to the average value of the losses decreases from 8/3 till $\pi/2$ with the increase of magnetic field.

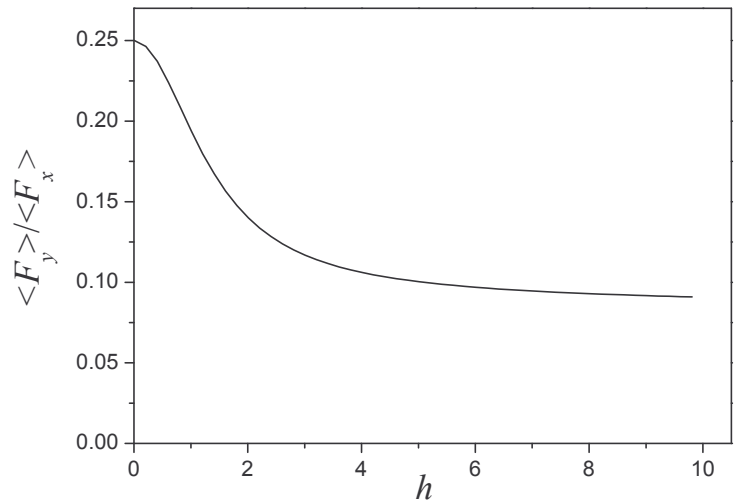

Fig. 4. Ratio of the average values of the functions $F_y$ and $F_x$ .

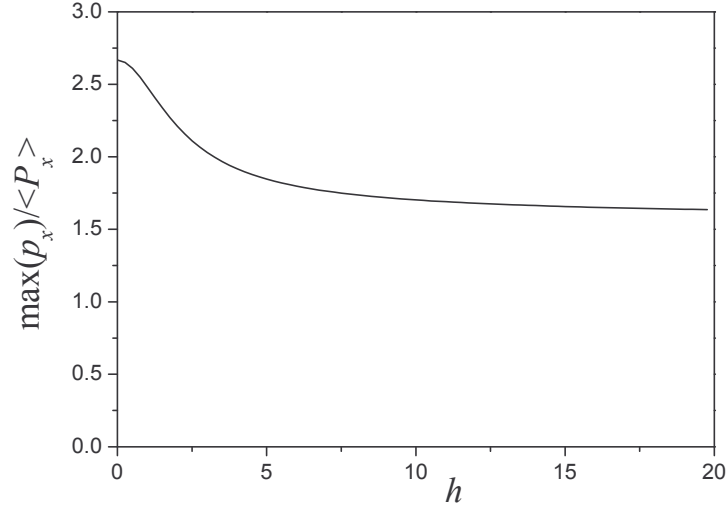

Fig. 5. Ratio of the loss maximum to the average losses as a function of the magnetic field.

At $h > h_l$, the approximation developed above becomes incorrect. The $y$-component of the current density attains the critical value $J_c$ and begins to influence on the distribution of the $x$-component. Simple expressions can be obtained only for the asymptotic case $h >> 1$. At $t = 0$, when the external field achieves its maximum, the sheet current density is equaled to $J_c$ almost at every point. The current distribution in the strip becomes as shown in Fig. 6. The current loops have a squared shape similar to the distributions obtained for thin films [12] and for bulk superconductors of finite sizes [13, 14]. The angular points are placed on lines at angle of $45^0$ to the $x$-axis. Note the difference of the distribution of the $y$-component of the current from that given by Eq. (6). For large $h$ the expression (6) gives narrow and high peaks of $J_{dy}$ near $kx = \pm\pi/2$. The peak height increases with $h$, and the width falls as $1/h$. As distinct from this behavior, in Fig. 6 $J_{dy}=J_c$ inside two triangles and zero in the rest area.

For $h >>1$ one can assume that at the decrease of the external field from $+H_0$ till $-H_0$ the directions of all the currents are changed by opposite in a negligible short time. Then the $x$-component of the electric field can be estimated as

$E_x (x, y, t) = -\omega\, \mu_0 H_0\, \sin(\omega t)\cos(kx)y$,

and the loss density is

$p_x = 2\mu_0\omega w^2 J_c H_c h\, \cos(kx)/\pi = P_0 h\cos(kx)$.     (16)

The loss density caused by the $y$-component of the current is calculated using (9) and (10):

$p_y \quad \dfrac{2\mu_0\omega H_c J_c kh}{3\pi}\left(w - L/4 + x\right)^3 \text{ for } L/4 - w \le x \le L/4$.     (17)

The similar expression for $p_y$ can be obtained for $-L/4 \le x \le -L/4 + w$. The maximums of these losses are

$$\max(p_x) = P_0 h, \tag{18}$$

$$\max(p_y) = P_0 h k w/3.$$

The average losses are

$$<P_x> = 2P_0\, h/\pi\,, \tag{19}$$

$$< P_y >= P_0 h \frac{(kw)^2}{6\pi}\,.$$

We obtain again that, at $L \gg w$, the total losses are determined mainly by the *x*-component. Fig. 7 shows how the losses calculated using Eq. (11) tend to their limits for $h \to \infty$.

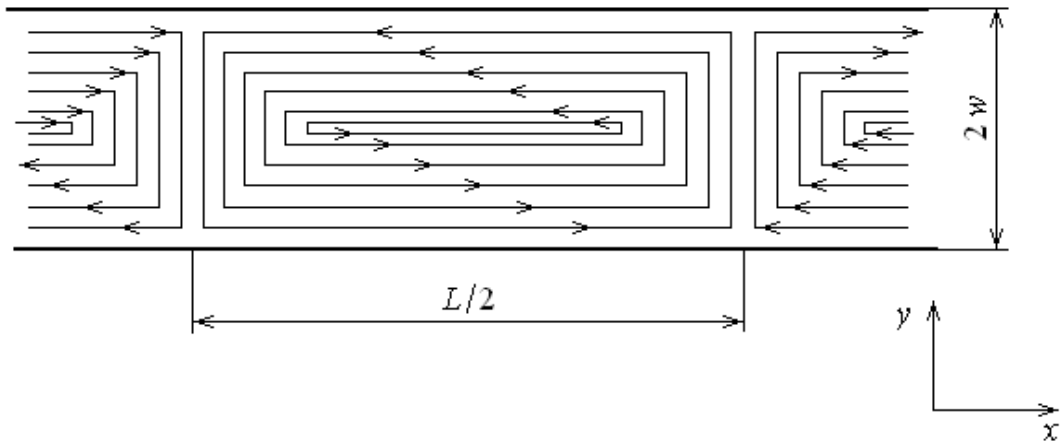

Fig. 6. Current distribution in a superconducting strip in the asymptotic case $h \gg 1$. Current density equals to the critical value at every point.

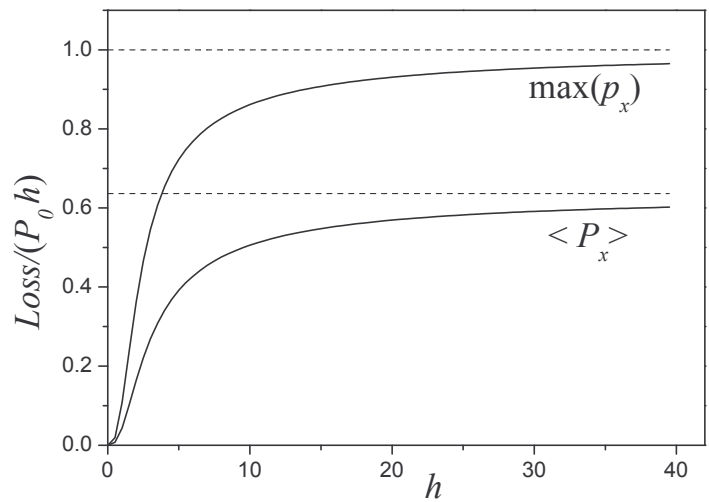

Fig. 7. Dependence of the average and maximum loss density $p_x$ on magnetic field. Dashed lines show the asymptotic values according to Eqs. (18) and (19).

## 3. Eddy current losses in a metal strip

### 3.1. A superconductor and a metal without an electrical contact between them

Let us first consider a superconductor-metal composite, where a superconducting strip is placed above a normal conducting strip with the same width, $2w$, and the thickness $a_m << w$ and there is no electrical contact between the superconductor and normal metal (Fig. 8). Such configuration can be considered for the simulation of coated superconductors where a superconducting layer is deposited on a dielectric buffer with negligible thickness.

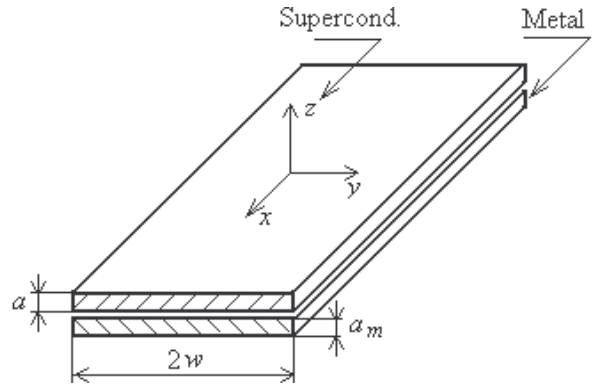

Fig. 8. Configuration of a superconducting-metal composite strip.

Losses in the metal strip will be calculated using in the low frequency limit assuming that the magnetic field generated by eddy currents in the metal can be neglected compared to an applied field. The magnetic field applied to the metal strip is the sum of the perpendicular external field $H = H_0\cos(kx)\cos(\omega t)$ and the field generated by currents flowing in the superconducting strip. Taking into account that $a, a_m << w$, the perpendicular component $H_z$ of the total field applied to the metal strip can be estimated as the magnetic field inside the superconducting strip. In a nonuniform field, two components of the current in the superconductor generate two parallel components ($x$ and $y$) of the magnetic field in the metal strip. The consideration analogous to the given above for a superconducting strip shows that, at $a_m << w$, the losses from the parallel fields in the metal strip can be negligible.

Analogously to the case of a superconducting strip we use the approximation $L >> w$ and determine the $x$-component of the induced current from the formulas for a uniform field where the field amplitude is replaced by the amplitude of a local field $H_a = H_0\cos(kx)$. Using the expressions obtained in [8] for the metal strip with the resistivity $\rho$ in a uniform perpendicular magnetic field, we can present the sheet current density $J_{mx}$ and loss density $p_{mx}$ in the form:

$$J_{mx}\left(x,y,t\right) = -\frac{a_m}{\rho}\frac{\partial\Phi}{\partial t} = \begin{cases} 0, & |y| < b_2 \\ \dfrac{\mu_0 a_m \sqrt{y^2 - b_2^2}}{\rho}\dfrac{y}{|y|}\dfrac{\partial H}{\partial t}, & b_2 < |y| < w \end{cases} \tag{20}$$

$$p_{mx} = \frac{8}{3\pi}\frac{\mu_0^2 a_m w^3 \omega^2 H_c^2}{\rho}F_m(Z) \tag{21}$$

where

$$F_m(Z) = \int_0^Z \sqrt{Z\eta - \eta^2}\left[1 - \frac{3}{\cosh(\eta)^2} + \frac{2}{\cosh(\eta)^3}\right]d\eta \tag{22}$$

and $Z = h\cos(kx)$.

The function $F_m$ is proportional to $h^6$ at $h \ll 1$ and to $h^2$ at $h \gg 1$. Detailed analysis of the dependence of the losses in a uniform field on the amplitude and frequency and their comparison with the hysteresis losses in a superconducting strip are presented in [8, 9].

The $y$-component of the current is estimated using Eq. (5). In contrast to the superconducting strip, in the metal strip the $x$-current component exists only at $|y| \geq b_2$. The $y$-component of the current exists for any $y$ and is independent of $y$ at $-b_2 < y < b_2$. The $y$-component of the current can be presented in the following form

$$J_{my}\left(x,y,t\right) = \frac{\mu_0 a_m \omega k w^2 H_c}{\rho}hF_i\left(x,y,h,t\right). \tag{23}$$

The function $F_i$ has a complex form. Let us estimate the ratio of the losses from the y- and x-components of the current. The loss density caused by the $y$-component of the current can be majorized:

$$p_{my} \leq \frac{\mu_0^2 a_m w^3 \omega^2 H_c^2}{\rho}\left(kw\right)^2 h^2 F_{imax}^2 \tag{24}$$

where the maximum value $F_{imax}$ of the function $F_i$ over $x$, $t$, and $y$ increases proportionally to $h^3$ at $h \ll 1$ and tends to 0.5 at $h \gg 1$.

It follows from Eqs. (21) and (24) that the ratio of both maximum and average values of $p_{my}$ and $p_{mx}$ is of order of $(kw)^2 h^2 \ll 1$ at $h \ll 1$ and $(kw)^2 \ll 1$ at $h \gg 1$. Thus, the losses caused by the $y$- component of the current in the metal strip can be neglected. The ratio of the maximum loss density to the average value for the $x$-component is given in Fig. 9. As for the case of a superconducting strip, the maximum is achieved at $x = 0$.

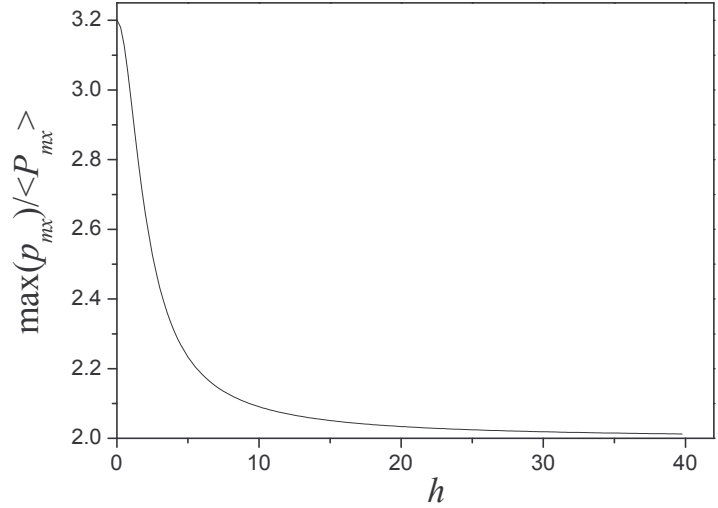

Fig. 9. Ratio of the maximum and average losses in the metal strip as a function of the external field.

As for a superconducting strip, obtained expressions for a metal strip are valid only under the condition (8), t. e. at $h < h_l$ . It is explained by the fact that the field applied to the metal is determined by the current distribution in the superconductor. At $h > h_l$ , let us consider the asymptotic case $h \gg 1$. The superconductor is in the critical state with the critical density $J_c$ and its contribution in the external field can be neglected. The $x$-components of the electric field and current induced in the metal strip are estimated as

$E_{mx} = \omega \, \mu_0 H_0 \cos(kx) \sin(\omega t) y$ ;

$J_{mx} = E_{mx} a_m / \rho = \mu_0 \omega H_0 \cos(kx) \sin(\omega t) y a_m / \rho$.

The loss density from the $x$-component is

$$p_{mx} = \mu_0^2 \omega^2 H_c^2 \cos(kx)^2 w^3 a_m h^2 / (3\rho) = P_{0m} h^2 \cos(kx)^2 \qquad (25)$$

where $P_{0m} = \mu_0^2 \omega^2 H_c^2 w^3 a_m / (3\rho)$.

The $y$-current component is

$$J_{my} = \int_{-w}^{y} \frac{\partial J_{mx}}{\partial x} \, dy = \frac{\omega k \mu_0 H_0 a_m \sin(kx) \sin(\omega t) \left( w^2 - y^2 \right)}{2\rho} \, .$$

The loss density from the $y$-component is

$$p_{my} = \frac{2}{15\rho} \omega^2 \mu_0^2 H_c^2 a_m w^3 h^2 \left( kw \right)^2 \sin(kx)^2 = \frac{2}{5} P_{0m} h^2 \left( kw \right)^2 \sin(kx)^2 \, . \qquad (26)$$

The ratio of the maximums and average values of the losses from $y$ and $x$ components becomes the same:

$$R = 2(kw)^2 / 5. \qquad (27)$$

We see that also in this case the losses from the *y*-component are negligible for $L \gg w$.

The loss density from the *x*-component calculated with Eqs. (21), (22) and their asymptotic (25) are shown in Fig. 10.

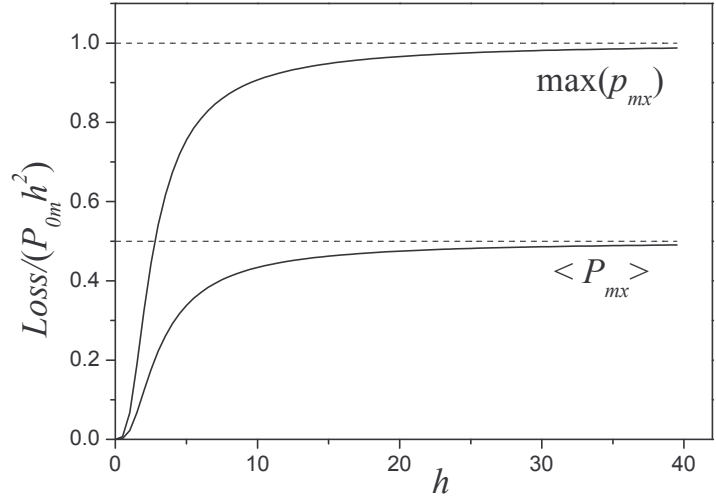

Fig. 10. Dependence of the average and maximum loss density $p_{mx}$ on magnetic field. Dashed lines show the asymptotic values according to Eq. (25).

It can be shown that in the framework of the used approximation the ratio between the losses from both current components remain of the same order also for the case when the width of the metal strip is larger than the width of the superconducting one.

The proportion between the losses in the superconductor and in the metal depends on the amplitude and frequency of the external field. These dependencies for a uniform field were investigated in details in [8] and they are approximately valid for the considered problem of a nonuniform field.

### 3.2. A superconductor and a metal in electrical contact

If superconducting and metal strips are in an electrical contact (tapes, coated superconductors, thin films covered by a metal layer), there is the possibility of the redistribution of the induced currents between the strips. Under the considered approximation, the *x*-components of the currents in the both strips are induced by a magnetic field while the *y*-components are closure currents appearing as a result of the non-uniformity of the field along *x*. The *x*-components of electric fields induced in the superconductor and metal are the same and exist where the *x*-component of the current in a superconductor achieves the critical value $J_c$, t. e. at $|y| > b_2$. In this region the superconductor and normal metal cannot shunt one another; there is no any redistribution of the current. The difference is observed in the region where $|J_x| < J_c$ and the $x$-component of the current in the metal $J_{mx} = 0$. At $h < h_l$, the $y$-component of the current induced in the superconductor is less than $J_c$ and the superconductor shunts the metal.

At $h > h_l$, when the condition (8) is violated, Eq. (6) gives a current density higher than $J_c$. In the case of an isolated superconductor, the increase of the external field leads to the extension of the region where the $y$-component achieves $J_c$ and tends to the distribution presented in Fig. 6. The electrical contact with the metal causes the competition between the extension of a zone of the critical state and the redistribution of the current into the metal, t. e. the metal shunts the superconductor.

In the case of the contact with a low-resistance metal the part of the $y$-component above $J_c$ flows in the metal and the total $y$-component in the composite can be estimated as

$$J_{ytot} = \int_{-w}^{y} \frac{\partial J_x}{\partial x} dy + \int_{-w_m}^{y} \frac{\partial J_{xm}}{\partial x} dy . \tag{28}$$

The second integral is taken over the width of the metal strip $w_m \geq w$. Losses can be evaluated from Eq. (9) with the substitution of $J_{ytot}$. In the considered case the limitation on the maximum magnetic field introduced by the condition (8) is canceled. The total sheet current density $J_{ytot}$ can exceed sufficiently the critical value. Therefore, in high magnetic fields, the maximum of a loss density from the $y$-component can be higher than the loss maximum from the $x$-component. As follows from Eq. (14), it can occur if $h > 40/(kw)^2$. In this case the maximum of the loss density is transferred to $kx \cong \pi/2$.

## 4. Applicability of the approximation

The approximate expressions for the loss calculation were obtained under a number of the assumptions. Let us analyze to which degree these assumptions correspond to practical cases. Typical tapes produced for application in power devices have the width of several millimeters, coated superconductors and films is about 1 cm. Therefore, the approach developed above is applicable for the AC fields spatially oscillating with the wave length $L$ longer than 20 cm. For example, in a power transformer the length $L$ is determined by the winding size and can be of the order of 1 m. The parameter $(kw)^2$ contained in the expressions for the losses from the $y$-component becomes 1/50 for $L = 20$ cm.

Another important parameter of the approximation is $H_c$ which is related to the sheet critical current density $J_c$. The value of $J_c$ is of the order of $10^3$ A/m for thin films, $10^4$ A/m for superconducting tapes [8, 9] and achieves $10^5$ A/m for a new type of coated superconductors [15]. Therefore $h = H_0/H_c$ is equal to 1 only for low magnetic fields ($\mu_0 H_0 \sim 4 \cdot 10^{-4}$-$4 \cdot 10^{-2}$ T). The region of the applicability of the approximation proposed for $h < h_l = \pi/(2kw)$ (condition (8)) can extend to fields $\mu_0 H_0 \sim 0.02$ - 2 T for $L \sim 1$ m. Note that the condition (8) is applicable for an isolated superconductor and loses its significance for the composite where a superconductor contacts with a well-conducting layer. In the last case the metal shunts the superconductor, and the y-component determined by Eq. (6) can be above $J_c$. The excess current flows in the metal and the maximum of the loss density from $y$-component can exceed the maximum from the x-component. It happens at $h > 40/(kw)^2$ that corresponds to the fields exceeding 2 T. This value exceeds possible spatial changes of AC field amplitude in real electric devices. Thus, in most practical cases it is enough to take into account only the losses generated by induced longitudinal currents in a composite strip. The relative contribution of closure currents decreases still more in case when a uniform AC field or a transport current is applied. This differs from the results known for superconducting cables where spatial changes of AC fields lead to sufficient increase of losses due to coupling currents (see [16, 17] and references in them). This discrepancy is explained by different configuration of composite conductors. In the cables, superconducting filaments are very thin with small losses, and the main part of the losses is determined by the losses in the normal-conducting matrix.

Figs. 7 and 10 show that the high-field asymptotic can be used for $h > 40$ ($\mu_0 H_0 \sim 0.02$ - 1.6 T). If $L = 20$ cm, $h_l = 10$ and the range remains where the obtained analytical solutions cannot be used. The losses in this range can be estimated as any value between the values given by two approaches: developed for $h < h_l$ and the high-field asymptotic. With increase $L$ this range is reduced and at large $L$ the regions of the applicability of two approaches can overlap.

All the expressions for losses were obtained above in the low frequency limit, t. e. under the assumption that the field generated by the eddy currents in the metal can be neglected compared to the external field. It is valid under the condition

$$\frac{\mu_0 w a_m \omega}{2\rho} \ll 1, \tag{29}$$

It was shown in [8, 9] that this assumption is valid till hundreds of kilohertz.

## 5. Conclusion

The analytical expressions were obtained for the evaluation of AC losses in superconductor-metal composite strips in a nonuniform AC magnetic field in the form of a standing wave.

It was shown that, at the calculation losses averaged over the length, the contribution of closure currents can be neglected compared to the part caused by longitudinal currents. However, at very high external magnetic fields (> 2 T), the maximum loss density from closure currents can be even above that one from longitudinal currents.

The obtained results can be used to evaluate losses in superconducting films, coated superconductors and monofilament tapes in spatially changing AC magnetic fields.

**Acknowledgements**

The authors would like to thank G. A. Levin for helpful discussions. This work was supported in part by the Ministry of Infrastructure of Israel.

**Appendix: Vector potential**

The external magnetic field is assumed to be $\vec{H} = H_0\cos(\omega t)[0,\ 0,\ \cos(kx)]$ at the surface of the superconductor, ($z \approx 0$). We assume that the field is independent of $y$ and near the plane $z = 0$ the external magnetic field can be presented as

$$\vec{H} \cong H_0\cos(\omega t)\left[\varphi_x(x,z),\ \varphi_y(x,z),\ \cos(kx)\left(1 + Cz^{\gamma}\right)\right] \qquad (A1)$$

where the functions $\varphi_x$ and $\varphi_y$ have zero values at $z = 0$, $C$ and $\gamma$ are constants that do not influence on the vector potential of the external field at $z = 0$ and $\gamma > 0$. The term $Cz^{\gamma}$ can be considered as the next one of a series.

From $\text{rot}\vec{H} = 0$ one can obtain $\varphi_y = 0$ and

$$\varphi_x = -k\sin(kx)\left( z + \frac{C}{\gamma+1}z^{\gamma+1} \right).$$

Assuming that the external field is induced by a current with zero $z$-component from

$$\text{rot}\vec{A} = \mu_0\vec{H} \qquad (A2)$$

it is following

$$A_x = \psi_x(x,y,t),$$

$$A_y = \mu_0 H_0 k\sin(kx)\cos(\omega t)\left[ \frac{z^2}{2} + \frac{Cz^{\gamma+2}}{(\gamma+1)(\gamma+2)} \right] + \psi_y(x,y,t). \qquad (A3)$$

The terms containing $z$ are not interesting for the future consideration since these terms vanish at loss calculation in the superconductor ($z = 0$). The functions $\psi_x$ and $\psi_y$ are determined by a solution of the following equations obtained using (A2) and $\operatorname{div}\vec{A} = 0$:

$$\frac{\partial \psi_y}{\partial x} - \frac{\partial \psi_x}{\partial y} = \mu_0 H_0 \cos(kx)\cos(\omega t), \tag{A5}$$

$$\frac{\partial \psi_x}{\partial x} + \frac{\partial \psi_y}{\partial y} = 0. \tag{A6}$$

Taking into account that the vector potential is determined accurate to gradient of any function, a periodic solution can be presented as

$$\psi_x = -\frac{\mu_0 H_0}{k}\cos(kx)\cos(\omega t)\sinh(ky), \tag{A7}$$

$$\psi_y = \frac{\mu_0 H_0}{k}\sin(kx)\cos(\omega t)\big[1 - \cosh(ky)\big]. \tag{A8}$$

The superconductor is placed at $|y| \le w$, $kw \ll 1$, and in a first approximation the vector potential at $z = 0$ is

$$\vec{A}_e = \mu_0 H_0 \cos(\omega t)\left[-y\cos(kx), \; -\frac{ky^2}{2}\sin(kx), \; 0\right].$$

**References**

[1] E. H. Brandt, M. Indenbom, Phys. Rev. **B 48** (1993) 12893.

[2] E. Zeldov, J. R. Clem, M. McElfresh, and M. Drawin, Phys. Rev. **B 49** (1994) 9802.

[3] W. T. Norris, J. Phys. D.: Appl. Phys. **4** (1971) 1358.

[4] W. J. Carr Jr., AC Loss and Macroscopic Theory of Superconductors, second ed., Teylor and Francis, New York, 2001.

[5] G. P. Mitikik , E. H. Brandt, and M. Indenbom, Phys. Rev. **B 70** (2004) 014520.

[6] E. Pardo, F. Gömöry, J. Šouc, and J. M. Ceballos, arXiv:cond-mat/050314, 12 Oct 2005.

[7] Y. Mawatari, Phys. Rev. **B 54** (1996) 13215.

[8] K.-H. Muller, Physica **C 281** (1997) 1.

[9] K.-H. Muller, Physica **C 312** (1999) 149.

[10] V. Sokolovsky, L. Prigozhin, and V. Meerovich, Physica C **408-410** (2004) 645.

[11] L. Prigozhin, and V. Sokolovsky, IEEE Trans. on Appl. Supercond. **14** (2004) 69.

[12] L. Prigozhin, J. Comput. Phys. **144** (1998) 183.

[13] A. Badía-Majós, and C. López, Appl. Phys. Lett. **86** (2005) 202510.

[14] A. Aydmer, and E. Yanmaz, Supercond. Sci. Technol. **18** (2005) 1010.

[15] S. R. Foltyn, H. Wang, L. Civale, Q. X. Jia, P. N. Arendt, B. Maiorov, and J. L. MacManus-Driscoll, Appl. Phys. Lett. **87** (2005) 162505.

[16] S. Takács, Supercond. Sci. Technol. **9** (1996) 137.

[17] S. Takács, IEEE Trans. on Appled. Supercond. **7** (1997) 258.